\documentclass[12pt,preprint]{aastex}

\slugcomment{Not to appear in Nonlearned J., 45.}
\shorttitle{Multi band optical follow-up of GRB020813}
\shortauthors{Urata et al.}
\begin{document}
\title{Multi-band optical follow-up observations of GRB 020813 at KISO and Bisei observatories}

\author{Y. Urata\altaffilmark{1,2}, S. Nishiura\altaffilmark{3},
T. Miyata\altaffilmark{3}, H. Mito\altaffilmark{3},
T. Kawabata\altaffilmark{4}, Y. Nakada\altaffilmark{3},
T. Aoki\altaffilmark{3}, T. Soyano\altaffilmark{3},
K. Tarusawa\altaffilmark{3}, A. Yoshida\altaffilmark{1,5},
T. Tamagawa\altaffilmark{1} and K. Makishima\altaffilmark{1,6}}

\altaffiltext{1}{RIKEN (Institute of Physical and Chemical Research), 2-1 Hirosawa, Wako, Saitama 351-0198, Japan}
\altaffiltext{2}{Department of Physics, Tokyo Institute of Technology, 2-12-1 Oookayama, Meguro-ku, Tokyo 152-8551, Japan}
\altaffiltext{3}{Kiso Observatory, Institute of Astronomy, The University of Tokyo, Mitake-mura, Kiso-gun, Nagano 397-0101, Japan}
\altaffiltext{4}{Bisei Astronomical Observatory, Ohkura, Bisei, Okayama 714-1411, Japan}
\altaffiltext{5}{Departments of Physics, Aoyama Gakuin University, 5-10-1 Fuchinobe, Sagamihara, Kanagawa 229-8558, Japan}
\altaffiltext{6}{Departments of Physics, The University of Tokyo, 7-3-1 Hongo, Bunkyo-ku, Tokyo 113-0033, Japan}

\begin{abstract}

Observations were made of the optical afterglow of GRB020813 (Fox,
Blake \& Price, 2002) with the KISO observatory 1.05 m Schmidt
telescope and the Bisei astronomical observatory 1.01 m
telescope. Four-band ($B, V, R$, and $I$) photometric data points were
obtained from 2002, August 13 10:52 to 16:46 UT, or 0.346$-$0.516 days
after the burst.  In order to investigate the early-time ($<$1 day)
evolution of the afterglow, four-band light curves were produced by
analyzing the data taken at these two astronomical observatories,
as well as publicly released data taken by the Magellan Baade
telescope (Gladders and Hall, 2002c).  The light curves can be
approximated by a broken power law, of which the indices are
approximately 0.46 and 1.33 before and after a break at $\sim$0.2
days, respectively.  The optical spectral index stayed approximately
constant at $\sim$0.9 over 0.17 $\sim$ 4.07 days after the burst.
Since the temporal decay index after the break and the spectral index
measured at that time are both consistent with those predicted by a
spherical expansion model, the early break is unlikely to be a jet
break, but likely to represent the end of an early bump in the light
curve as was observed in the optical afterglow of GRB021004.

\end{abstract}


\section{Introduction}

Afterglow of a gamma-ray burst (GRB), observed in X-ray through radio
frequencies, can be interpreted in the fireball models, wherein a
shock produced by the interaction of relativistic ejecta with the
circumburst environment expands into the surrounding medium, producing
broad-band synchrotron emission (Meszaros \& Rees 1997; Sari, Piran,
\& Narayan 1998).  If the GRB is collimated into a jet, the entire jet
surface becomes visible to the observers at some time $t_{j}$. As the
jet starts to expand laterally at around $t_{j}$ (Rhoads 1999), its
sweeping area increases faster than before, leading to a stronger
deceleration and hence a faster afterglow decay.  The rapid decay of
some GRB afterglows, observed recently, provide evidence for such
jet-like or collimated ejecta (Sari, Piran, \& Halpern 1999). The jet
model would relax the energy requirements on some of the more extreme
GRBs by a factor of several hundred (Frail et al. 2001).

The bright and long event, GRB020813, was detected on 2002 August 13,
with the {\it HETE-2} spacecraft (Ricker et al. 2003).  The flight
localization was reported in a GCN Position Notice at 02:48:33 UT, 4
min 14 s after the burst trigger.  The subsequent ground analysis of
the {\it HETE-2} data produced a refined burst location, which
was reported in a GCN Position Notice at 05:48:35 UT, 184 minutes
after the burst.  The location, with 90\%-confidence error radius of
$60''$ (due entirely to systematic errors), is centered at
$\alpha^{2000} = 19^{\rm h}46^{\rm m}38^{\rm s}, \delta^{2000} = -
19^{\circ} 35' 16''$ (Villasenor et al. 2002).

The optical afterglow was found within the $60''$-radius error circle
at 0.078 days (112 minutes) after the burst, at the coordinates of
$\alpha^{2000}= 19^{\rm h}46^{\rm m}41.^{\rm s}88,
\delta^{2000}=-19^{\circ}36'05''.1$ (Fox et al., 2002).  Optical
spectra of the afterglow taken with the Keck observatory
exhibit numerous absorption lines, indicating a minimum redshift of
1.254 $\pm$ 0.005 (Price et al. 2002).  Early optical light curves of
the afterglow are suggested to have undergone a temporal break at
3.5$-$5 hours after the burst (Bloom et al. 2002).

The Kiso observatory of the University of Tokyo and the Bisei
Astronomical Observatory (BAO) have established, for the first time in
Japan, the capability of multi-band follow-up observations of GRBs
(Urata et al. 2003a). The two sites serve as valuable additions to the
world-wide optical and infrared follow-up network, because the Japan
area would otherwise be blank for the network. 

In order to investigate the early (within 1 day of the burst)
evolution of the optical afterglow of GRB020813 in the flux
(e.g. temporal decay and existence of a jet break) and the spectral
slope, we have analyzed the Kiso and BAO data. In addition, we analyze
the publicly released data of the afterglow obtained by the Magellan
Baade telescope (Gladders and Hall, 2002c).

\section{Observations}

We carried out follow-up observations of the optical afterglow of
GRB020813 with the 1.05 m Schmidt telescope and a 2k$\times$2k CCD
Camera at the Kiso observatory, starting at 2002 August 13 10:52 UT
(0.339 days after the burst). The field of view is $51.'2\times51.'2$
and the pixel size is $1''.5$ square.  We performed $B$-, $V$-, $R$-,
and $I$-band observations using a system prepared for the GRBs
follow-up observations (Urata et al. 2003a). We obtained the
multi-band data as described in talbe 1, with each dataset consisting
of triple frames.

We also performed $R$-band observations using the 1.01-m telescope
with a Mutoh CV16IIE CCD camera (Kodak KAF1602E chip) at the BAO,
starting at 2002 August 13 11:44:19 UT (0.375 days after the
burst). The field of view is $7'.8\times5.'2$ and the pixel size is
$0''.9$ square.  We obtained 36 frames of $R$-band data, each with 60
s exposure (table 1).

An extensive observation of the afterglow was also performed with the
Magellan Baade 6.5 m telescope, on the nights of August 13 and 14
(0.13 - 0.94 days after the burst). These data were obtained in the
$B$-, $V$-, $R$-, and $I$-band each for 60 s, using a TEK5 camera
(Gladders and Hall,2002c). The four-band images, after bias
subtraction and flat-fielding, are made publicly available by Gladders
and Halls (2002c). We have hence retrieved the images from the ftp
site introduced by them. The data consist of 4, 13, 14, and 21 frames
in the  $B$-, $V$-, $R$-, and $I$-band, respectively.

\section{Analysis and Results}
\subsection{Photometry}
We processed the Kiso and BAO data by a standard method using the NOAO
IRAF.  We used appropriate calibration data for the bias-subtraction
and flat-fielding corrections. An example of the $I$-band images we
obtained is shown in figure 1.  Thus, the afterglow is clearly
detected in the images of all observations listed in table 1.

The Kiso, BAO, and Baade data have been combined with median.
Flux calibrations among the different sites were done using the APPHOT
package in IRAF, referring to the standard stars suggested by Henden
(2002).  Specifically, we utilized three of them, at the J2000
coordinates of ($19^{\rm h}46^{\rm m}41^{\rm s}.1672$, $-19^{\circ}
36' 00''.454$), ($19^{\rm h}46^{\rm m}43^{\rm s}.9488$, $-19^{\circ}
36' 01''.166$), and ($19^{\rm h}46^{\rm m}44^{\rm s}.5416$,
$-19^{\circ} 35' 41''.662$).  We set the aperture size to 4 time as
large as the FWHM of objects for each data.  We summarize the results
of our photometry in table 1.

\subsection{Light curves}

Figure 2 shows four-band light curves of the optical afterglow based
on our measurements.  We thus cover a time period of 0.17 to 0.94 days
after the burst in the $V$-, $R$-, $I$-band, with the Kiso, BAO and Baade
datasets.  The Baade datasets densely cover the early phase ($<$ 0.2
days), with an additional coverage at 0.94 days. The Kiso and BAO
data, in contrast, constrain the light curves in the intermediate
(0.35 - 0.52 days) range.

First, we tried to fit the $V$-, $R$-, and $I$-band light curves by a
simple power law of the form $\propto t^{-\alpha}$, where $t$ is the
time after the burst onset and $\alpha$ is a constant called decay
index. This gave $\alpha=1.05$ with reduced $\chi^{2}(\chi^{2}/\nu)$
of 2.44 for the $V$-band, $\alpha=1.03$ with $\chi^{2}/\nu=$ 4.12 for
the $R$-band, and $\alpha=0.89$ with $\chi^{2}/\nu=$ 53.6 for the
$I$-band.  Thus, none of the three light curves are consistent with a
single power-law decay.

Next, we tried to fit the light curves with a broken power law model
expressed as
\begin{eqnarray}
F_{t} = f_{*}(t/t_{*})^{-\alpha1}[1 - \exp(-J)]/J;\\
J(t,t_{*},\alpha1,\alpha2) = (t/t_{*})^{(\alpha2 - \alpha1)}, \nonumber
\end{eqnarray}
where $f_{*}$, $t_{*}$, $\alpha_{1}$, and $\alpha_{2}$ are four
parameters.  This functional form has no physical significance, but it
provides a good description to the GRB990510 data, with the asymptotic
power-law indices being $\alpha_{1}$ and $\alpha_{2}$ at early and
late times, respectively (Harrison et al. 1999).

We have successfully fitted the above function to the $V$-, $R$-, and
$I$-band light curves. For the $V-$band, we have obtained
$\alpha_{1}=0.46\pm0.05$, $\alpha_{2}=1.35\pm0.01$, and
$t_{*}=0.22\pm0.01$ ($\chi^{2}/\nu=0.60$ with $\nu=$ 11); for the
$R-$band, $\alpha_{1}=0.46\pm0.06$, $\alpha_{2}=1.33\pm0.01$, and
$t_{*}=0.21\pm0.01$ ($\chi^{2}/\nu=0.92$ with $\nu=$12); for the
$I-$band, $\alpha_{1}=0.30\pm0.02$, $\alpha_{2}=1.30\pm0.01$, and
$t_{*}=0.23\pm0.01$ ($\chi^{2}/\nu= 1.19$ with $\nu=$17).
Thus, the decay is relatively independent of color, as evidenced by
nearly the same values of $t_{*}\sim 0.2$ days found among the three
bands.  An initial analysis of the optical light curve suggested
$t_{j} \sim 3.5-5$ hours (Bloom et al. 2002; Gladders \& Hall 2002b).

We fitted a single power law to the $B$-band data, because we lack
data before the break. The obtained index is $\alpha = 1.32\pm0.03$,
with $\chi^{2}/\nu=$0.26. Thus, the slope agrees with the values of
$\alpha_{2}$ found in the three longer wavelengths.

The X-ray afterglow, observed by the {\it Chandra} HETG at 0.88 - 1.78
days after the burst, faded in brightness according to a power law,
with a decay index of $1.42\pm0.05$ (Vanderspek et al. 2002). This is
close to the optical values we measured after the break.

\subsection{Spectral flux distributions}

We have converted the $BVRI$ magnitudes to fluxes using the effective
wavelengths and normalizations of Fukugita et al (1995).  To remove
the effects of the Galactic interstellar extinction, we used the
reddening map of Schlegel, Finkbeiner, \& Davis (1998).  The Galactic
reddening toward the burst is significant, $E(B-V$)=0.101, which
implies a Galactic extinction of $A_{B}=0.44$, $A_{V}=0.34$,
$A_{R}=0.27$, and $A_{I}=0.20$.

In figure 3, we plot the spectral flux distributions obtained in this
way using the Kiso data (0.43 days after the burst) and the Baade data
(0.17, 0.20 and 0.94 days after the burst). We fitted them with a
power-law function as $f(\nu)\propto \nu^{-\beta}$, where $f(\nu)$ is
the flux density at frequency $\nu$ and $\beta$ is the spectral index.
We have obtained $\beta = 0.93 \pm 0.04$, $0.82 \pm 0.03$, $0.93 \pm
0.16$, and $0.91 \pm 0.07$ at $t$= 0.17, 0.20, 0.43, and 0.94 d,
respectively. The former two values of $\beta$ were calculated using
the $V$-, $R$-, and $I$-band data, while the latter two values
utilized the four-band data. Thus, the measured values of $\beta$ are
consistent with being constant at $\beta=0.87\pm0.03$, because fitting
them with a constant yields $\chi^{2}/\nu=1.82$ with $\nu=$3.

The optical spectral index of the GRB020813 afterglow was reported as
$\beta=1.06\pm0.01$ from red side Keck spectra
(5600$-$9400$\mathaccent 23A$) taken at about $t$= 0.19 d (Barth et
al. 2003).  Correcting it for the Galactic extinction quoted above, we
obtain $\beta \sim0.7$.
Levan et al. (2002) reported $\beta=0.8$ from the {\it HST} imaging
data take at $t$= 4.08 d. Since these measurements consistently imply
$\beta$=0.7$\sim$0.9, the spectral slope $\beta$ is suggested to have
remained rather constant over a period of $t$ = 0.17 to 4.08 d.

\section{Discussion}
We analyzed the GRB020813 afterglow images in four bands, observed
over a period of 0.33$\sim$0.52 days after the burst at the Kiso and
Bisei observatories. We also re-analyzed the publicly released
four-band images taken by the Baade 6.5m telescope (Gladders \& Hall
2003c).
These datasets, when combined, define relatively well-sampled light
curves (in $B$-, $V$-, $R$-, and $I$-band) covering $t$= 0.17 to 0.94
d. The $V$-, $R$-, and $I$-band light curves are rather similar, and
can be described by a broken power-law of $\alpha_{1} \sim 0.4$ and
$\alpha_{2} \sim 1.3$, with a break point at $t\sim 0.2$ d.
Meanwhile, the spectral index of the afterglow remained at $\beta
\sim$0.9. As can be seen from figure 2, the present Kiso and BAO data
are essential in the determination of the overall light curves.

Optical light curves of several GRBs exhibit a temporal break at 
 $t\sim$ 1 d or later. The decay index of
the light curve is typically $\sim 1$ and $\sim 2$, before and after
the break, respectively.  In contrast, the afterglow of GRB020813
showed an unusually early break time at 0.2 d  So
far, the earliest break was observed in GRB010222, at 0.73 d
(Watanabe et al. 2001). In addition, the decay index of $\alpha_{1}
\sim 0.46$ observed from the present afterglow is smaller than those
of other afterglows, while $\alpha_{2}$ is similar to $\alpha_{1}$ of
other bursts.  These are reminiscent of the afterglow of GRB021004;
its light curve showed an early bump characterized by a brightening
phase lasting until $\sim$0.07 d, followed by a dimming with a
decay index of $\sim 0.2$ (Urata et al. 2003b) which is close to the
present $\alpha_{1}$.

Usually, these temporal breaks are successfully interpreted by the jet
model (Rhoads 1999; Sari, Piran, \& Halpern 1999).  The jet model
predicts the time dependent spectral flux to vary as
$F_{\nu}(t)\propto \nu^{-\beta}t^{-\alpha}$, wherein the temporal
decay index $\alpha$ and the spectral index $\beta$ are both
determined uniquely by the electron power law spectral index $p$.  In
order to examine whether the jet model is applicable to the present
afterglow, we utilized the relations by Sari et al. (1999), and
calculated the values of $p$ and $\alpha$ based on the observed
$\beta$.  The results are shown in table 2, together with the employed
relations. Thus, the vales of $\alpha_1 = 0.3-0.46$ we observed are
too flat to be reproduced by the jet model under any condition, while
those of $\alpha_2 = 1.30-1.35$ agree with the model prediction
assuming that we are observing a spherical fireball (i.e., {\em before
the break}) in the frequency range below the synchrotron cooling.

The above discussion suggests that the behavior of the present light
curves {\em after} their apparent break at 0.2 d is consistent with
what is expected {\em before} the jet break in terms of the standard
jet model.  Therefore, the observed break is probably distinct from
the usual jet break.  If so, the electron spectral index is inferred
from table 4 to be $p=2.8$, which falls at the steepest end of the
distribution of $p$ (1.4--2.8; Panaiteschu and Kumar, 2001); it is
close to that of GRB980519 ($p \sim 2.8$).  If the jet break exists in
this burst, it should occur later than $t$=0.94 d.  We suggest that
the mild break observed in the present light curves is reminiscent of
a similar break, which was observed in the decay phase of an early
bump in the GRB021004 afterglow.  The GRB021004 afterglow showed a
brightening phase over 0.05$\sim$0.07 d after the burst based on the
Kiso observation, followed by a temporal break around 0.2 d across
which the decay index changed from 0.2 to 0.7 (Urata et al. 2003b).
In the case of GRB021004, the optical color changed over the bump, but
remained constant across the break like in the present light curves;
the behavior may be explained by a crossing of the typical synchrotron
frequency through the optical band (Kobayashi and Zhang, 2003).

\acknowledgments
 The authors thank Gladders \& Hall for their observations as well as
for making their data publicly available.  Y. Urata acknowledges
support from the Japan Society for the Promotion of Science (JSPS)
through JSPS Research Fellowships for Young Scientists.

\clearpage
\begin{figure}
\plotone{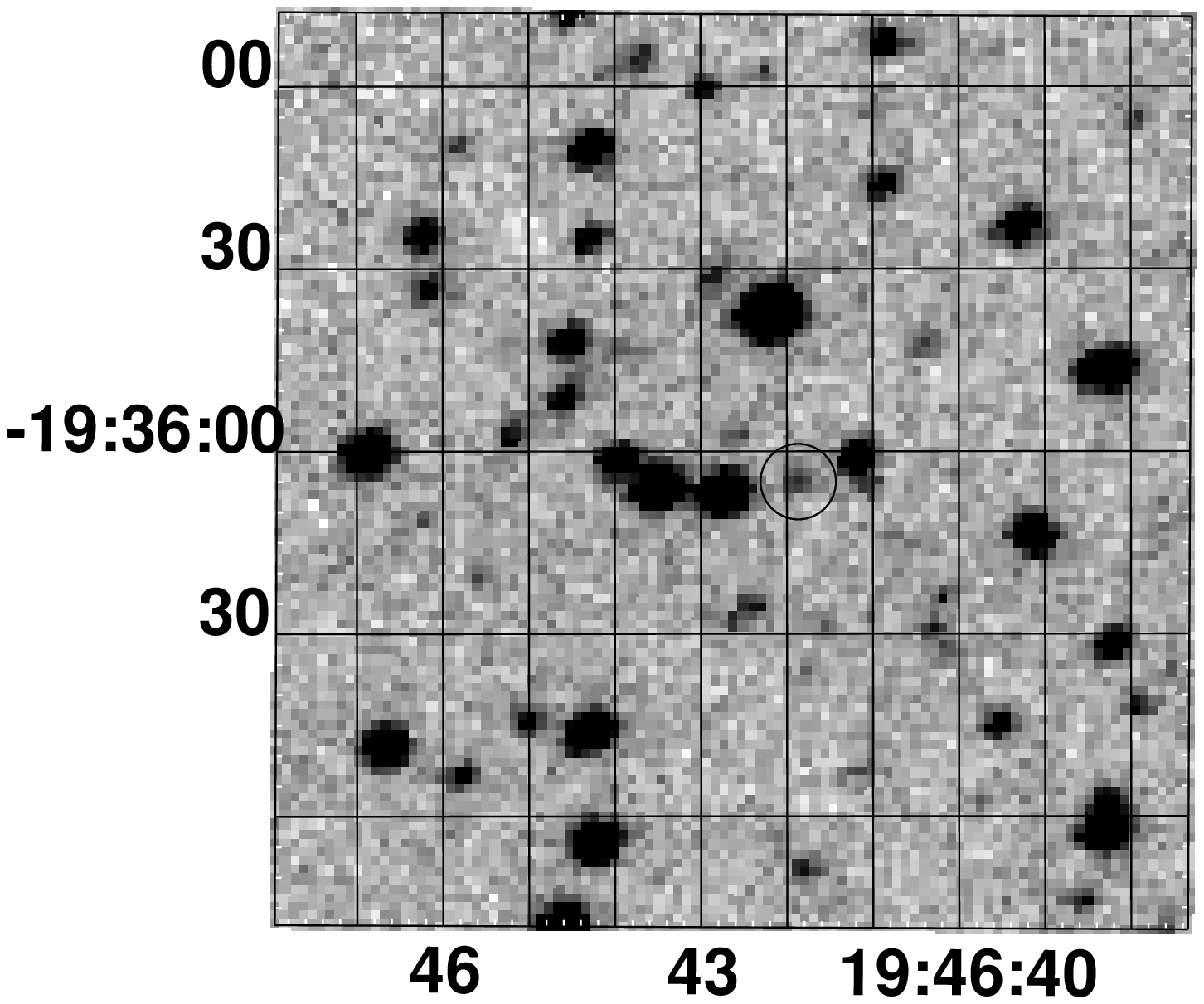}
\caption{An $I$-band image of the GRB020813 field obtained at the Kiso
observatory, with a 300-s exposure starting at 2002 August 13 11:14
UT. The afterglow is indicated by a circle near the image center.\label{fig1}}
\end{figure}

\begin{figure}
\plotone{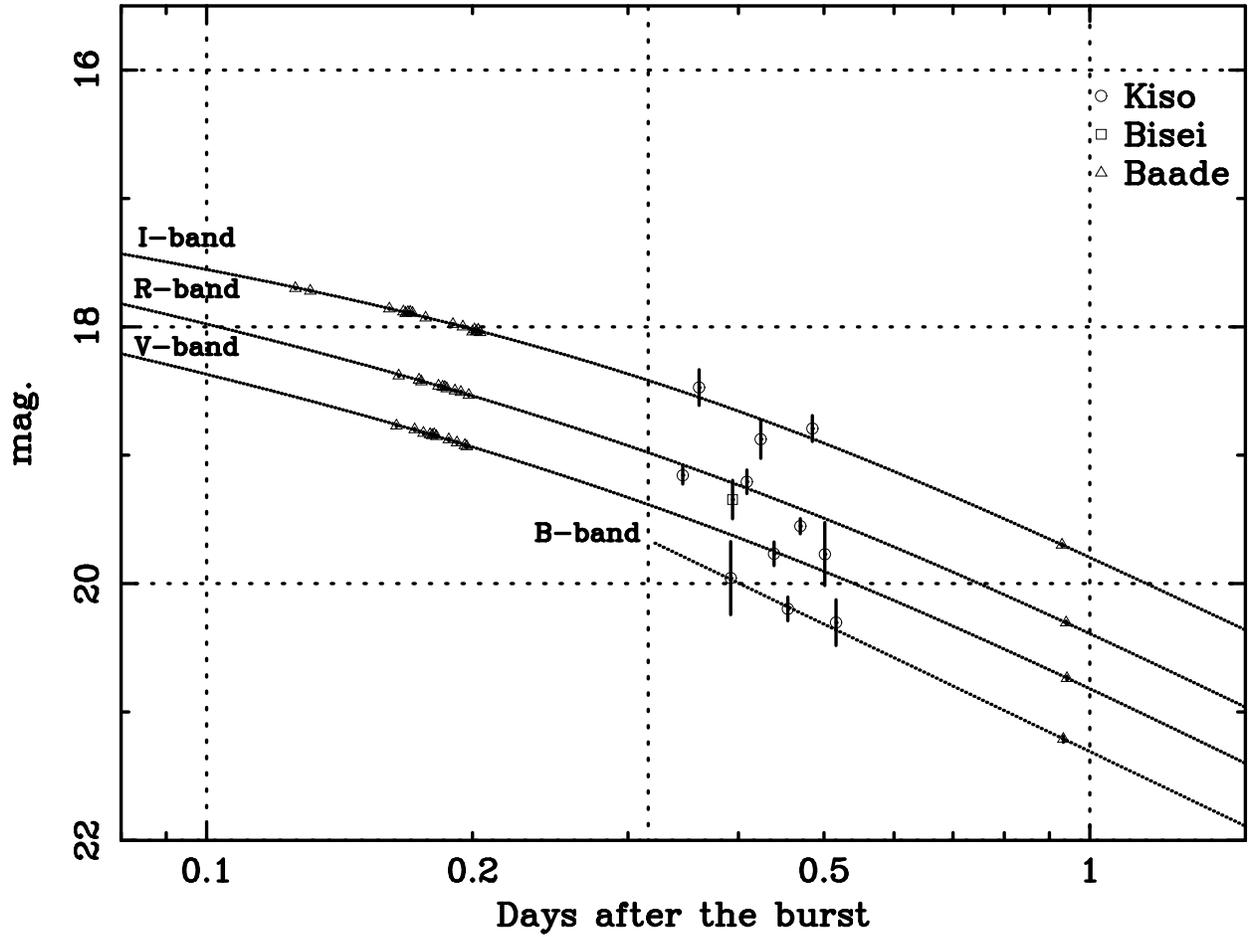}
\caption{The $B$-, $V$-, $R$-, and $I$-band light curves based on the photometry of Kiso,
Bisei and Baade. The smooth curves represent the broken power law model (Harrison et
al. 1999) fitted to the $V$-, $R$-, and $I$-band light curves.\label{fig2}}
\end{figure}

\begin{figure}
\plotone{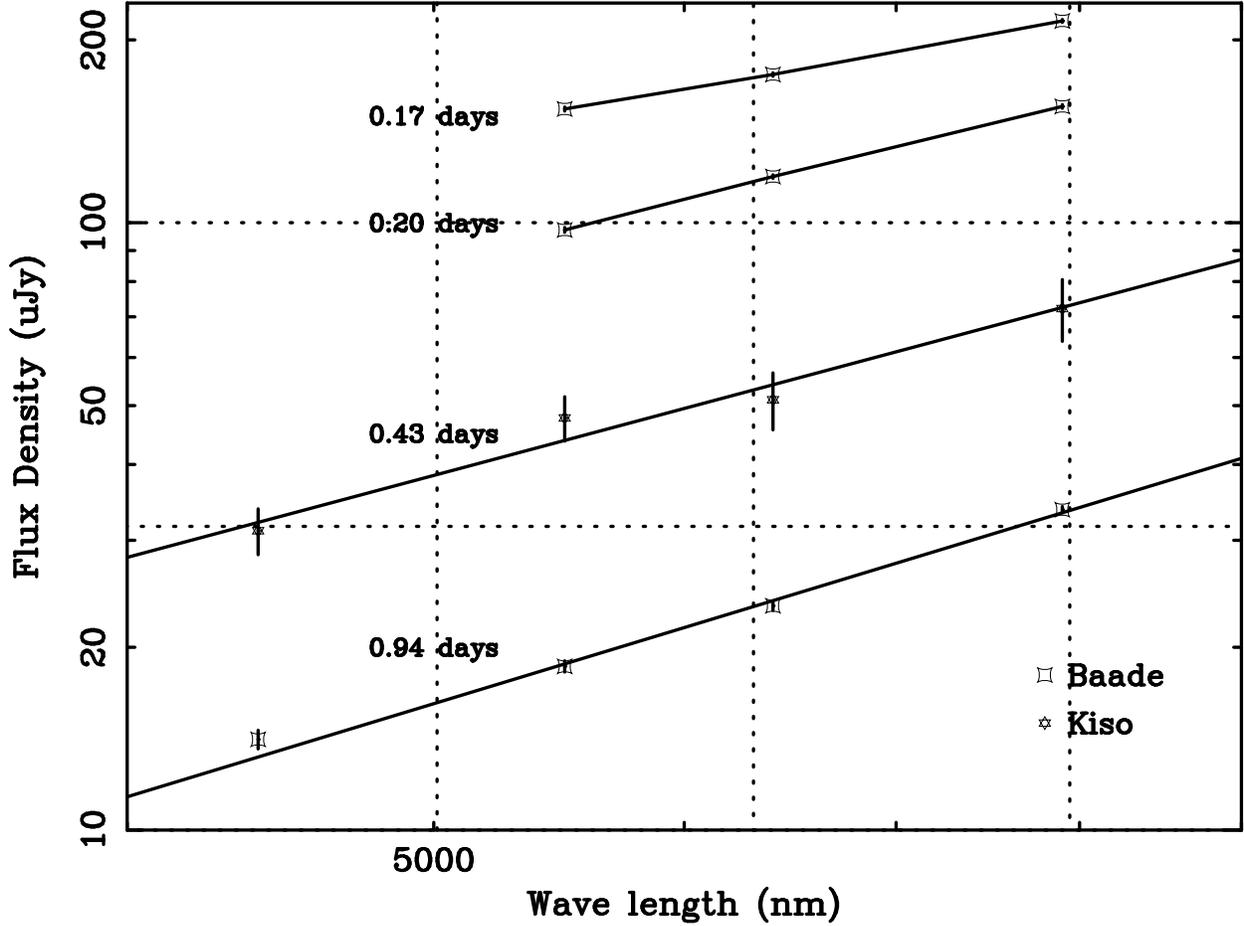}
\caption{The spectral flux distributions of the optical afterglow
associated with GRB020813, determined with the Kiso data (0.43 d)
and the Baade data (0.17, 0.20, and 0.94 d). The straight lines
give the best-fit power-law function described in the
text\label{fig3}}
\end{figure}

\clearpage 

\begin{table}
\begin{center}
\caption{Log of Kiso and Bisei follow-up observations made on 2002 August 13.\label{tbl-1}}
\begin{tabular}{ccccc}
\tableline\tableline
Start Time(UT) & Filter & Exposure (s) & mag & Site\\
\tableline
11:58:04 & $B$ & 300 s $\times$  3 & 19.96 $\pm$ 0.29 & Kiso\\
13:29:06 & $B$ & 300 s $\times$  3 & 20.20 $\pm$ 0.09 & Kiso\\
14:57:03 & $B$ & 300 s $\times$  3 & 20.30 $\pm$ 0.18 & Kiso\\ 
13:07:08 & $V$ & 300 s $\times$  3 & 19.77 $\pm$ 0.09 & Kiso\\
14:35:12 & $V$ & 300 s $\times$  3 & 19.77 $\pm$ 0.25 & Kiso\\
10:52:26 & $R$ & 300 s $\times$  3 & 19.16 $\pm$ 0.07 & Kiso\\
11:44:19 & $R$ &  60 s $\times$ 36 & 19.35 $\pm$ 0.13 & Bisei\\
12:23:20 & $R$ & 300 s $\times$  3 & 19.21 $\pm$ 0.09 & Kiso\\
13:51:16 & $R$ & 300 s $\times$  3 & 19.55 $\pm$ 0.06 & Kiso\\
11:14:23 & $I$ & 300 s $\times$  3 & 18.47 $\pm$ 0.14 & Kiso\\
12:45:17 & $I$ & 300 s $\times$  3 & 18.87 $\pm$ 0.15 & Kiso\\
14:13:19 & $I$ & 300 s $\times$  3 & 18.79 $\pm$ 0.10 & Kiso\\
\tableline
\end{tabular}
\end{center}
\end{table}

\begin{table} 
\begin{center} 
\caption{Values of the decay index $\alpha$ and the electron spectral index $p$,
calculated from the measured spectral index $\beta=0.91\pm0.07$.\label{tbl-4}}
\begin{tabular}{ccccc} 
\tableline\tableline 
Frequency\tablenotemark{*} & phase & relation\tablenotemark{\dagger} & $\alpha$\tablenotemark{\ddagger} & $p$ \\ 
\tableline 
$\nu_{c} > \nu_{\rm opt}$  & sphere & $\alpha=3\beta/2, p=2\beta+1$   & $1.37\pm0.11$ & $2.81\pm0.14$\\
$\nu_{c} > \nu_{\rm opt}$  & jet    & $\alpha=2\beta+1, p=2\beta+1$   & $2.82\pm0.14$ & $2.81\pm0.14$\\
$\nu_{c} < \nu_{\rm opt}$  & sphere & $\alpha=3\beta/2-1/2, p=2\beta$ & $0.87\pm0.11$ & $1.81\pm0.14$\\
$\nu_{c} < \nu_{\rm opt}$  & jet    & $\alpha=2\beta, p=2\beta$       & $1.82\pm0.14$ & $1.81\pm0.14$\\
\tableline 
\tablenotetext{*}{$\nu_{\rm c}$ is the frequency at which the spectrum
breaks due to synchrotron cooling, whereas $\nu_{\rm opt}$ is the
typical visible light frequency.}  
\tablenotetext{\dagger}{The jet model relation due to Sari et al. (1999).}
\tablenotetext{\ddagger}{The decay index calculated from the spectral
index $\beta$.}
\end{tabular} 
\end{center} 
\end{table}


\begin{thebibliography}{}
\bibitem[Barth et al., (2003)]{keck} Barth, A. J. et al., 2003, \apj, 584, 47
\bibitem[Bloom et al.(2002)]{break} Bloom, J. S., Fox, D. W., and Hunt, M. P., 2002, GCN Circ. 1476
\bibitem[Fox, Blake \& Price, GCN1470]{OTfind} Fox, D. W., Blake, C., and Price, P., 2002, GCN Circ. 1470
\bibitem[Frail et al., (2001)]{frail} Frail, D. A., et al., 2001, \apj, 562, L55
\bibitem[Fukugita et al., (1995)]{fukugita} Fukugita, M., Shimasaku, K., Ichikawa, T., 1995, \pasp, 107, 945
\bibitem[Gladders1(2002)]{Phot} Gladders, M. and Hall, P., 2002a, GCN Circ. 1495
\bibitem[Gladders2(2002)]{Decayindex} Gladders, M. and Hall, P., 2002b, GCN Circ. 1514
\bibitem[Gladders3(2002)]{Baade} Gladders, M. and Hall, P., 2002c, GCN Circ. 1519
\bibitem[Harrison et al., (1999)]{Harrison} Harrison, F. A. et al., 1999, \apj, 523, L121 
\bibitem[Henden(2002)]{Henden} Henden, A., 2002, GCN Circ. 1503
\bibitem[Kawabata et al.,(2002)]{Bisei} Kawabata, T., Urata, Y., and Yamaoka, H., 2002, GCN Circ. 1501
\bibitem[Kobayashi and Zhang. (2003)]{kobayashi} Kobayashi, S., and Zhang, B., 2003, \apj, 582, L75
\bibitem[Levan et al., (2002)]{hst} Levan, A. J., Fruchter, A. S., Burud, I., and Rhoads, J. E., 2002, GCN Circ. 1518  
\bibitem[Meszaros and Rees (1997)]{mes} Meszaros, P. and Rees, M. J., 1997, \apj, 482, L29 
\bibitem[Panaitescu et al.,(2001)]{beta} Panaitescu, A. and Kumar, P. 2001, \apj, 560, L49\
\bibitem[Pian et al., (2001)]{SAX} Pian, E., et al., 2001, A\&A, 372, 456 
\bibitem[Piran et al., (1999)]{model1} Piran, T. 1999, Phys. Rep., 314, 575 
\bibitem[Price et al.,(2002)]{redshift} Price, P., Bloom, J. S., Goodrich, R. W., Barth, A. J., Cohen, M. H. and Fox, D. W., 2002, GCN Circ. 1475
\bibitem[Rhoads (1999)]{rhoads}  Rhoads, J. E., 1999, \apj, 525, 737 
\bibitem[Ricker et al., 2003]{hete} Ricker, G.R. et al., 2003, AIP conf. proce., 662, 3
\bibitem[Sari et al.,(1999)]{sari} Sari, R., Piran, T. and Halpern, J. P., 1999, \apj, 519, L17
\bibitem[Sari et al., (1998)]{sari2}  Sari, R., Piran, T. and Narayan, R., 1998, \apj, 497, L17 
\bibitem[Schlegel et al., (1999)]{map} Schlegel, D. J., Finkbeiner, D. P. and Davis, M., 1998, \apj, 500, 525 
\bibitem[Stanek et al., (1999)]{stanek}  Stanek, K. Z., Garnavich, P. M., Kaluzny, J., Pych, W., and Thompson, I., 1999, \apj, 522, L39
\bibitem[Urata1 et al.,(2002)]{KISO} Urata, Y. et al., 2002, GCN Circ. 1485
\bibitem[Urata et al.,(2003)]{kisosystem} Urata, Y., Nakada, Y., Miyata, T., Nishiura, S., Mito, H., Aoki, T., Soyano, T., Tarusawa, K., 2003a, Roma2002 proceedings 
\bibitem[Urata et al., (2003)]{GRB021004} Urata, Y. et al., 2003b, ICRC2003 proceedings
\bibitem[Vanderspek et al., (2002)]{chandra} Vanderspek, R., Marshall, H. L., Ford, P. G., and Ricker, R., 2002, GCN Circ. 1504
\bibitem[Villasenor et al., (2002)]{gamma} Villasenor, J. et al., 2002, GCN Circ. 1471
\bibitem[Watanabe et al., (2001)]{watanabe} Watanabe, J., Kinoshita, D., Komiyama, Y., Fuse, T., Urata, Y., and Yoshida, F., 2001, PASJ, 53, L27

\end{thebibliography}
\end{document}